\documentclass[]{spie}  

\usepackage{amsmath,amsfonts,amssymb}
\usepackage{subcaption}
\usepackage{graphicx}
\usepackage{cite} 
\usepackage{epsfig}
\usepackage{nccmath}
\usepackage{xcolor,colortbl}
\usepackage{tabularx}
\usepackage{relsize}
\usepackage{pifont}
\usepackage{booktabs} 
\usepackage{multirow}
\usepackage{multicol}
\usepackage{adjustbox}
\usepackage{float}
\usepackage{makecell}
\usepackage{tabu}
\usepackage[colorlinks=true, allcolors=blue]{hyperref}
\usepackage[capitalize]{cleveref}
\usepackage{algorithm}
\usepackage{adjustbox}

\title{Evaluating New AI Cell Foundation Models on Challenging
Kidney Pathology Cases Unaddressed by Previous Foundation Models}

\author[a$\dag$]{Runchen Wang}
\author[b$\dag$]{Junlin Guo}
\author[b]{Siqi Lu}
\author[c]{Ruining Deng}
\author[b]{Zhengyi Lu}
\author[a]{Yanfan Zhu}
\author[a]{Yuechen Yang}
\author[b]{Chongyu Qu}
\author[d]{Yu Wang}
\author[d]{Shilin Zhao}
\author[a,b]{Catie Chang}
\author[b]{Mitchell Wilkes}
\author[e]{Mengmeng Yin}
\author[e]{Haichun Yang}
\author[a,b,e]{Yuankai Huo}

\affil[a]{Department of Computer Science, Vanderbilt University, Nashville, TN 37235, USA}
\affil[b]{Department of Electrical and Computer Engineering, Vanderbilt University, Nashville, TN 37235, USA}
\affil[c]{Department of Radiology, Weill Cornell Medicine, New York, NY 10021, USA}
\affil[d]{Department of Biostatistics, Vanderbilt University Medical Center, Nashville, TN 37235, USA}
\affil[e]{Department of Pathology, Microbiology and Immunology, Vanderbilt University Medical Center, Nashville, TN 37235, USA}

\affil[$\dag$]{These authors contribute equally to this work.}

\authorinfo{Further author information: (Send correspondence to Yuankai Huo)\\
E-mail: yuankai.huo@vanderbilt.edu}

\begin{document}
\maketitle

\begin{abstract}
Accurate cell nuclei segmentation is critical for downstream tasks in kidney pathology and remains a major challenge due to the morphological diversity and imaging variability of renal tissues. While our prior work has evaluated early-generation AI cell foundation models in this domain, the effectiveness of recent cell foundation models remains unclear. In this study, we benchmark advanced AI cell foundation models (2025), including CellViT++ variants and Cellpose-SAM, against three widely used cell foundation models developed prior to 2024, using a diverse large-scale set of kidney image patches within a human-in-the-loop rating framework. We further performed fusion-based ensemble evaluation and model agreement analysis to assess the segmentation capabilities of the different models. Our results show that CellViT++ [Virchow] yields the highest standalone performance with 40.3\% of predictions rated as ``Good'' on a curated set of 2,091 challenging samples, outperforming all prior models. In addition, our fused model achieves 62.2\% ``Good'' predictions and only 0.4\% ``Bad", substantially reducing segmentation errors. Notably, the fusion model (2025) successfully resolved the majority of challenging cases that remained unaddressed in our previous study. These findings demonstrate the potential of AI cell foundation model development in renal pathology and provide a curated dataset of challenging samples to support future kidney-specific model refinement.

\end{abstract}

\section{Introduction}
Nuclei instance segmentation is a critical task in computational pathology~\cite{deng2025casc, chen2025systematic}, serving as a foundational step for downstream analyses such as cell counting~\cite{ke2023clusterseg, cui2024pfps}, phenotype classification~\cite{pratapa2021image,yang2025pyspatial}, and spatial transcriptomics~\cite{zhu2025asign, zhu2025magnet}. However, traditional deep learning approaches often depend on extensive pixel-level annotations and struggle to generalize across various tissue types, staining protocols, and imaging modalities. This limits their scalability and clinical applicability, especially in high-variance domains such as kidney pathology. These challenges have motivated a shift to foundation models, large-scale pre-trained architectures built for broad adaptability across diverse biomedical tasks~\cite{cui2024pfps, ma2024segment, li2023llava}.

In our prior work (2024) ~\cite{guo2025assessment}, we evaluated three widely used cell foundation models—Cellpose~\cite{cellpose}, StarDist~\cite{stardist}, and CellViT~\cite{cellvit}—on a large-scale kidney pathology dataset consisting of 8,789 high-resolution image patches sampled from 2,542 kidney whole slide images (WSIs). Although these models demonstrated promising zero-shot performance, they still consistently failed in challenging image patches, such as those with low contrast, overlapping nuclei, or atypical morphologies. As large-scale deep learning continues to evolve, a key question arises: \textbf{Can newly released cell foundation models overcome the limitations of previous models?} 

\begin{figure}[ht]
    \centering
    \includegraphics[width=1\textwidth]{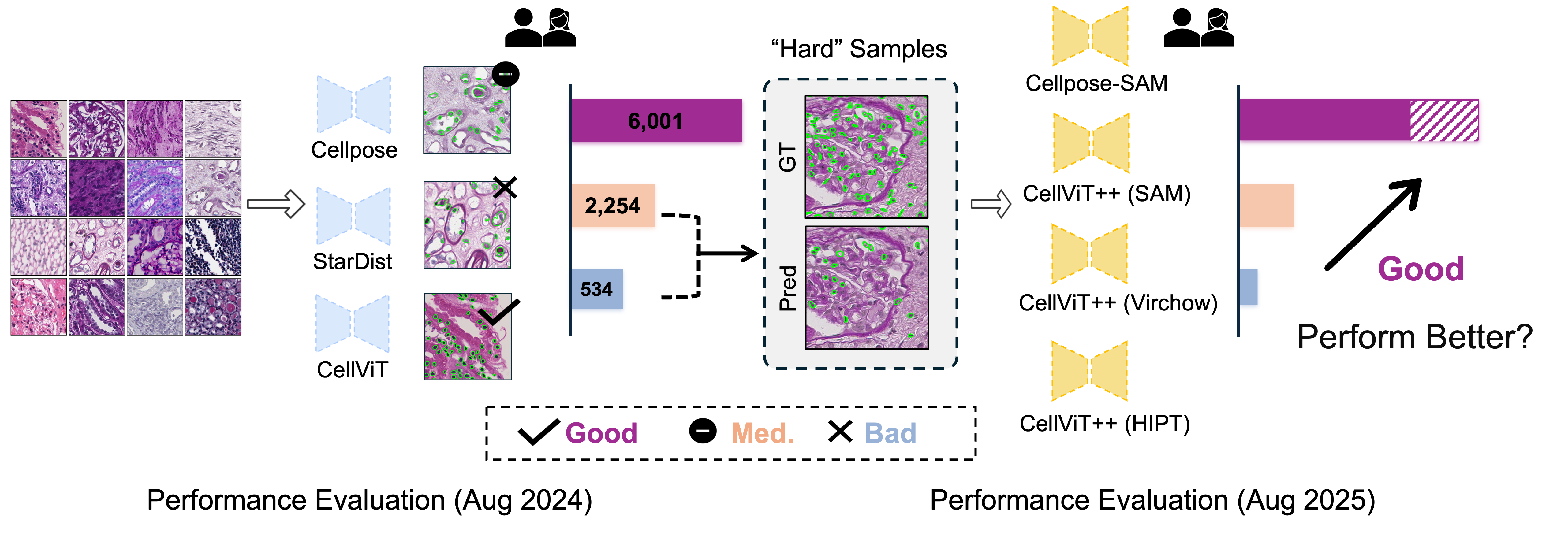}
    \caption{\textbf{Overall framework.} The previous study (Aug 2024) performed kidney cell nuclei instance segmentation using three cell foundation models: Cellpose~\cite{cellpose}, StarDist~\cite{stardist}, and CellViT~\cite{cellvit}. It revealed segmentation limitations in hard kidney pathology cases. These difficult cases were re-evaluated and categorized in Aug 2025 using four state-of-the-art models: three CellViT++~\cite{horst2025cellvit++} variants and Cellpose-SAM~\cite{pachitariu2025cellpose}. Model performance was evaluated based on human-in-the-loop rating for each prediction mask.}
    \label{fig:fig1}
\end{figure}

Motivated by this question, we conduct a comprehensive evaluation of newly released (2025) AI cell foundation models for histopathology, focusing on the task of nuclei instance segmentation in the kidney. These include CellViT++~\cite{horst2025cellvit++} and Cellpose-SAM~\cite{pachitariu2025cellpose}, which leverage large-scale pretrained Vision Transformers (e.g., SAM~\cite{kirillov2023segany}, Virchow~\cite{vorontsov2024virchow}, HIPT~\cite{chen2022scaling}) and are integrated into established cell instance segmentation frameworks. This enables better generalization and segmentation accuracy, particularly in the challenging or ambiguous samples identified in our prior assessment. A standardized inference pipeline is used across models and human-in-the-loop quality ratings are incorporated to assess segmentation quality in a realistic setting, as in our prior work. The overall framework for this study is shown in Fig.~\ref{fig:fig1}. 

\noindent {In summary, our study makes the following contributions:}  
\begin{itemize}  
    \item \textbf{Benchmarking new AI cell foundation models}: We provide the first comprehensive evaluation of three CellViT++ variants (HIPT, SAM, Virchow) and Cellpose-SAM on challenging kidney pathology patches that prior models failed to segment, highlighting their evolving capabilities.  

    \item \textbf{Human-in-the-loop evaluation of hard cases}: We extend our assessment to 2,091 curated difficult samples, rated into 
    “Good”,“Medium”,“Bad” categories, ensuring clinically meaningful performance measurement.  

    \item \textbf{Demonstrated robustness and precision}: By targeting hard cases missed in our 2024 study, we show that newer models achieve stronger generalization and higher-quality segmentation.  

\end{itemize}

\section{Methods}

\begin{figure}[!h]
    \centering
    \includegraphics[width=1\textwidth]{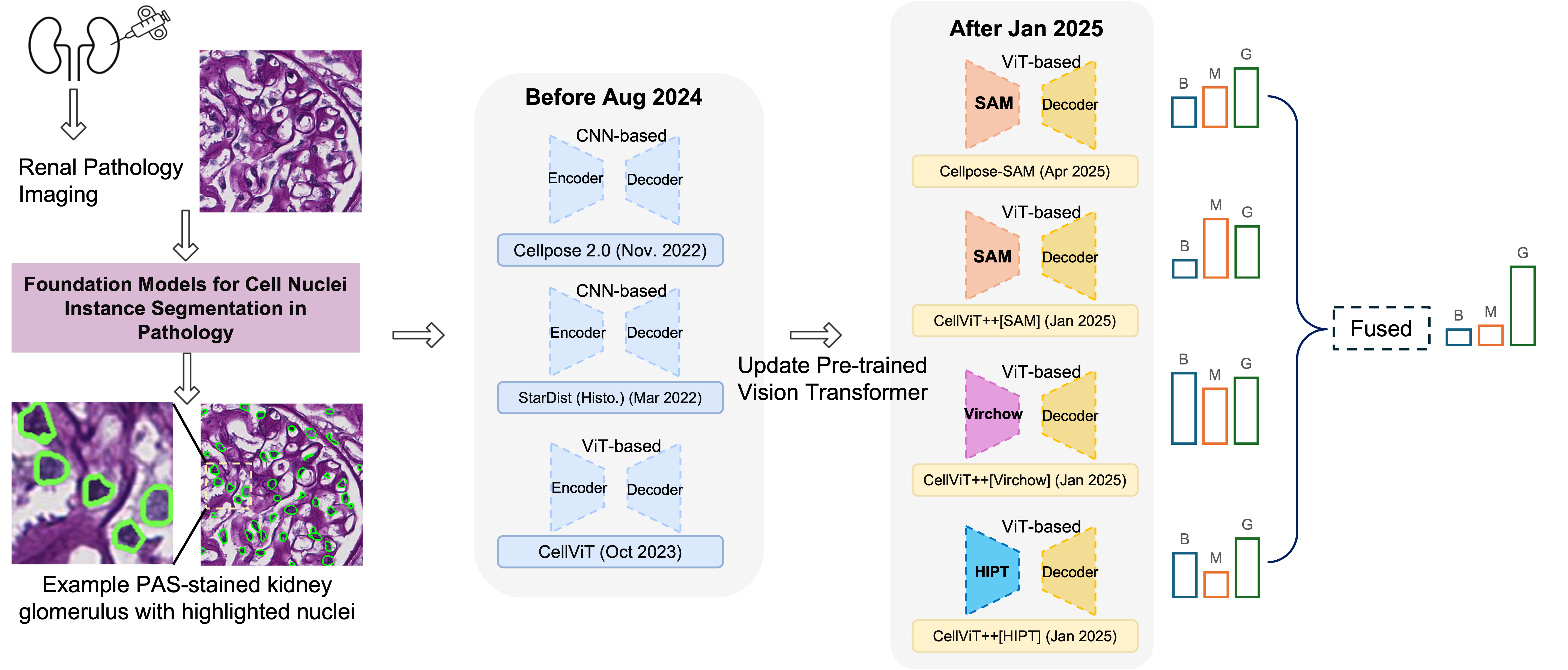}
    \caption{\textbf{Illustration of new AI cell foundation models in this evaluation.} Compared to our earlier work (before August 2024, shown in blue), the new models adopt vision transformer architectures for both encoder and decoder. These encoders (shown in distinct colors) incorporate large-scale pretrained foundation models such as HIPT~\cite{chen2022scaling}, Virchow~\cite{vorontsov2024virchow}, SAM~\cite{kirillov2023segany}.}
    \label{fig:evaluate_models}
\end{figure}




Fig.~\ref{fig:evaluate_models} presents the new AI cell foundation models evaluated in this study. On the left, the cell nuclei instance segmentation task is shown. Compared to our earlier work (before August 2024, shown in blue), the new models adopt vision transformer architectures for both encoder and decoder. These encoders (shown in distinct colors) incorporate large-scale pretrained foundation models such as HIPT~\cite{chen2022scaling}, Virchow~\cite{vorontsov2024virchow}, SAM~\cite{kirillov2023segany}.


\subsection{Evaluated Cell Nuclei Foundation Models}
In this assessment, we benchmark four state-of-the-art cell nuclei vision foundation models on challenging or uncertain samples that were unaddressed in our prior studies.~\cite{guo2025assessment, guo2024Good}. Details of the models are summarized below and in Table \ref{tab:table1}.

\begin{enumerate}
    \item \textbf{CellViT++[HIPT/SAM/Virchow]}: CellViT++~\cite{horst2025cellvit++} is a generalized cell segmentation framework for H\&E-stained images. Compared to its predecessor, CellViT~\cite{cellvit}, it leverages large-scale, next-generation vision foundation model pretrained weights (e.g., HIPT~\cite{chen2022scaling}, Virchow~\cite{vorontsov2024virchow}, SAM~\cite{kirillov2023segany}). Similar to CellViT, CellViT++ is fine-tuned for pathology cell nuclei segmentation using the PanNuke~\cite{gamper2019pannuke} dataset.

    \item \textbf{Cellpose-SAM}: Cellpose-SAM~\cite{pachitariu2025cellpose} integrates the pretrained transformer backbone from the SAM \cite{kirillov2023segany} foundation model into the Cellpose framework. Like the original Cellpose~\cite{cellpose}, which uses a U-shaped CNN backbone, it outputs horizontal and vertical topological gradients along with semantic predictions. Instance segmentation is derived using the Cellpose gradient flow tracking algorithm.
\end{enumerate}


\begin{table}[ht]
\centering
\caption{State-of-the-Art Cell Nuclei Segmentation Models in Pathology (2025)}
\label{tab:cell-foundation-models}
\begin{adjustbox}{max width=\textwidth}
\begin{tabular}{@{}lllc@{}}
\toprule
\textbf{Model}             & \textbf{Release Date} & \textbf{Backbone}          & \textbf{Post-processing}                                  \\ \midrule
CellViT++ (SAM)~\cite{horst2025cellvit++}            & 2025 Jan              & ViT (SAM-H)~\cite{kirillov2023segany}                      & HoVer-Net~\cite{hover-net}                               \\
CellViT++ (HIPT)~\cite{horst2025cellvit++}           & 2025 Jan              & ViT (HIPT\textsubscript{256})~\cite{chen2022scaling}                   & HoVer-Net~\cite{hover-net}                               \\
CellViT++ (Virchow)~\cite{horst2025cellvit++}        & 2025 Jan              & ViT (Virchow)~\cite{vorontsov2024virchow}                    &       HoVer-Net~\cite{hover-net}                         \\
Cellpose-SAM~\cite{pachitariu2025cellpose}               & 2025 Apr              & ViT (SAM)~\cite{kirillov2023segany}          & Gradient Tracking~\cite{cellpose}    \\
\bottomrule
\end{tabular}
\end{adjustbox}
\label{tab:table1}
\end{table}

\subsection{Rating Criteria}
\begin{figure}[!h]
    \centering
    \includegraphics[width=0.9\textwidth]{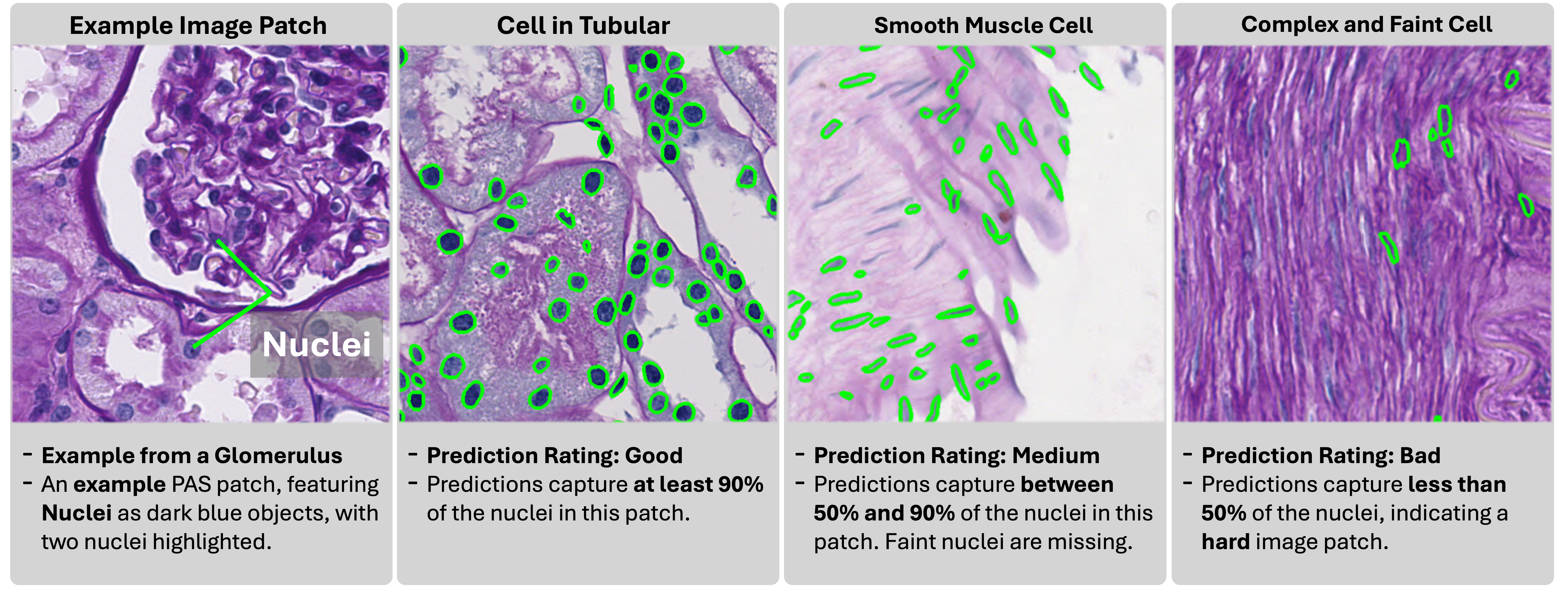}
    \caption{\textbf{Rating criteria for this assessment.} Illustrative details of the different rating category examples are provided.}
    \label{fig:criteria}
\end{figure}

 To assess segmentation quality, we applied rating criteria from our prior studies~\cite{guo2025assessment, guo2024Good}, developed by a renal pathologist with over 10 years of experience. As shown in Fig.~\ref{fig:criteria}, two trained students reviewed each model’s predictions and assigned one of three labels: ``Good'', ``Medium'', or ``Bad.''
This framework ensured consistent, quantitative evaluation. The curated patches also serve as valuable resources for training or fine-tuning future foundation models in kidney pathology.

\section{Data and Experiments}

\textbf{Implementations:} The model inference pipeline was implemented using Python~3.13.5 and PyTorch~2.7.1, with GPU acceleration provided by CUDA~12.9. All experiments were conducted on an NVIDIA RTX A5000 GPU with 24\,GB of memory. The rating was conducted using the Python GUI toolkit, Tkinter.

\subsection{Evaluation Dataset} 
In this study, the evaluation dataset is sourced from our previous assessment, comprising 8,789 image patches sampled from 2,542 WSIs stained with H\&E, PAS, or PASM. These WSIs were obtained from public datasets (e.g., KPMP~\cite{kpmp_data}, NEPTUNE~\cite{barisoni2013digital}, HuBMAP~\cite{hubmap-kidney-segmentation}) and in-house data from Vanderbilt University Medical Center. All images were cropped into 512 $\times$ 512-pixel patches at 40$\times$ magnification.

This second-round assessment focuses on difficult samples unaddressed in the prior studies~\cite{guo2025assessment, guo2024Good}. A curated subset of 2,091 patches—rated as ``Medium'' or ``Bad'' and exhibiting greater segmentation uncertainty—was selected as the evaluation dataset. 

\subsection{Individual and Fused Model Performance}

We systematically evaluated the performance of each foundation model across our evaluation dataset and further examined a fused strategy that ensembled their outputs. The individual model evaluation focused on the distribution of different prediction classes (“Good”, “Medium”, “Bad”), providing insights into the performance and behavior of each model. The prediction class assignments were determined through the rating criteria in Fig.~\ref{fig:criteria}.

Building on each cell foundation model’s performance, we also constructed a fused model based on the agreement of model-generated ratings. A patch was labeled as “Good” if at least one model's output was rated as "Good", and “Bad” only if all models were rated as "Bad". Patches not meeting either condition were labeled as “Medium.” This fusion scheme simulates the potential benefit of model complementarity and also serves to identify common failure cases—patches rated as "Bad" across all models highlight challenging cases and direct future model improvement.


\subsection{Cross-Model Performance Evaluation}

Similar to our prior work~\cite{guo2025assessment}, we conducted a cross-model agreement analysis to assess consistency and divergence among the four cell foundation models. We first computed a \textbf{pairwise agreement matrix}, which captures the percentage of image patches where each model pair assigns the same quality label (``Good", ``Medium", or ``Bad"). This quantifies alignment between model predictions and reveals systematic differences in behavior. We further categorized each patch based on consensus levels across all four models—\textbf{All Agree, Three Agree, Two Agree, and No Agreement}—to highlight class-wise consistency and guide ensemble strategy design.

\subsection{Comparison with Previous Study}

To enable a rigorous comparison with our prior 2024 study, we compared the performance of the updated 2025 fusion model with the 2024 fusion results across two evaluation scenarios: (1) a challenging subset of 2,091 image patches curated for hard cases, and (2) the full set of approximately 8,789 image patches from the 2024 evaluation. For consistency, the same rating protocol and annotation schema (``Good", ``Medium", ``Bad") were applied. This comparative setup ensures that observed performance differences reflect genuine model improvements rather than variations in evaluation design or data distribution.

\section{Results}
\subsection{Individual Model Performance Evaluation.}
\begin{figure}[!h]
    \centering
    \includegraphics[width=1.0\textwidth]{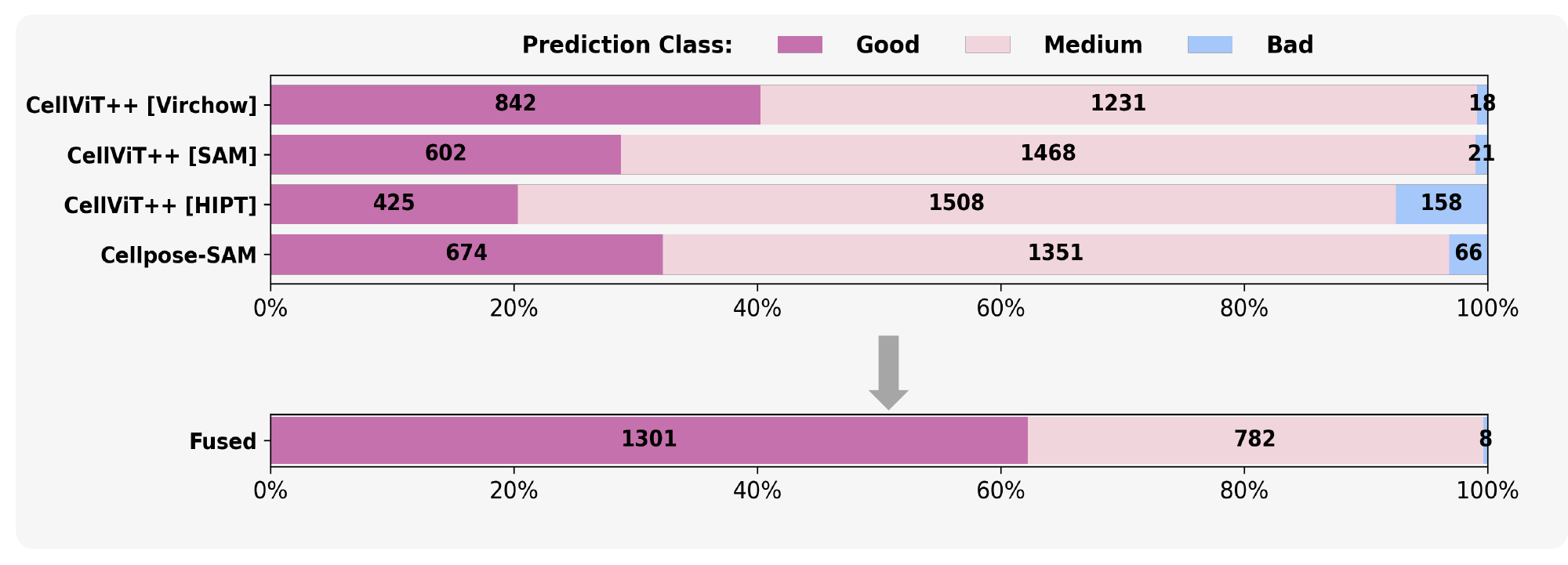}
    \caption{\textbf{Distribution of rated predictions from AI cell foundation models across the evaluation dataset.} Each row represents the foundation model’s predictions, with three values corresponding to the number of predictions rated as “Good,” “Medium,” and “Bad,” respectively. Then, data enrichment (shown as “Fused” Model) was performed based on the evaluation results of individual models, resulting in an increase in “Good” image patches and a decrease in “Bad” image patches.}
    \label{fig:individual}
\end{figure}

\noindent We began by evaluating the standalone performance of each foundation model on the task of cell nuclei instance segmentation on kidney pathology image patches using human ratings. The upper panel of Fig.~\ref{fig:individual} presents the distribution of predictions categorized as ``Good", ``Medium", or ``Bad" based on segmentation quality across four models: CellViT++ [HIPT], CellViT++ [SAM], CellViT++ [Virchow], and Cellpose-SAM.

Among the models, CellViT++ [Virchow] achieved the strongest performance, with 842 patches (40.3\%) rated as ``Good" and only 18 (0.9\%) as ``Bad", reflecting high generalizability to kidney pathology and suggesting a high degree of alignment between its predictions and human judgment.  Cellpose-SAM and CellViT++ [SAM] followed closely, with 674 (32.2\%) and 602 (28.8\%) ``Good" ratings, respectively, and fewer than 2.9\% ``Bad" predictions. In contrast, CellViT++ [HIPT] produced the proportion of ``Good" ratings (425, 20.3\%) and the highest ``Bad" ratio (158, 7.6\%), indicating weaker robustness on this evaluation dataset in comparison to other models. Although the majority of predictions for all models were rated as "Medium" (58.9--72.1\%), the differences in ``Good" and ``Bad" outcomes underscore the improvement of instance segmentation quality from new AI cell foundation models. These improvements reflect the impact of large-scale pretraining on model generalization for complex cell nuclei in pathology.

\subsection{Fused Model Performance Evaluation.}

To evaluate whether combining outputs from multiple foundation models enhances segmentation reliability, we implemented a fused model that aggregates their ratings. As shown in the bottom panel of Fig.~\ref{fig:individual}, the fused model achieved 62.2\% ``Good" ($n=1,301$) and 37.4\% ``Medium" ($n=781$) predictions, with only 0.4\% ($n=8$) rated as ``Bad". This represents a substantial reduction in segmentation failures compared to any individual model, suggesting that the fusion strategy effectively mitigates individual model weaknesses.

Notably, this improvement was achieved without retraining or additional supervision, demonstrating the efficacy of ensemble integration in leveraging complementary model strengths. The near-elimination of ``Bad" predictions highlights the potential of fusion-based strategies to deliver more robust and dependable cell segmentation in complex kidney pathology settings.

\subsection{Cross-Model Performance Agreement. }
\begin{figure}[!h]
    \centering
    \includegraphics[width=1.0\textwidth]{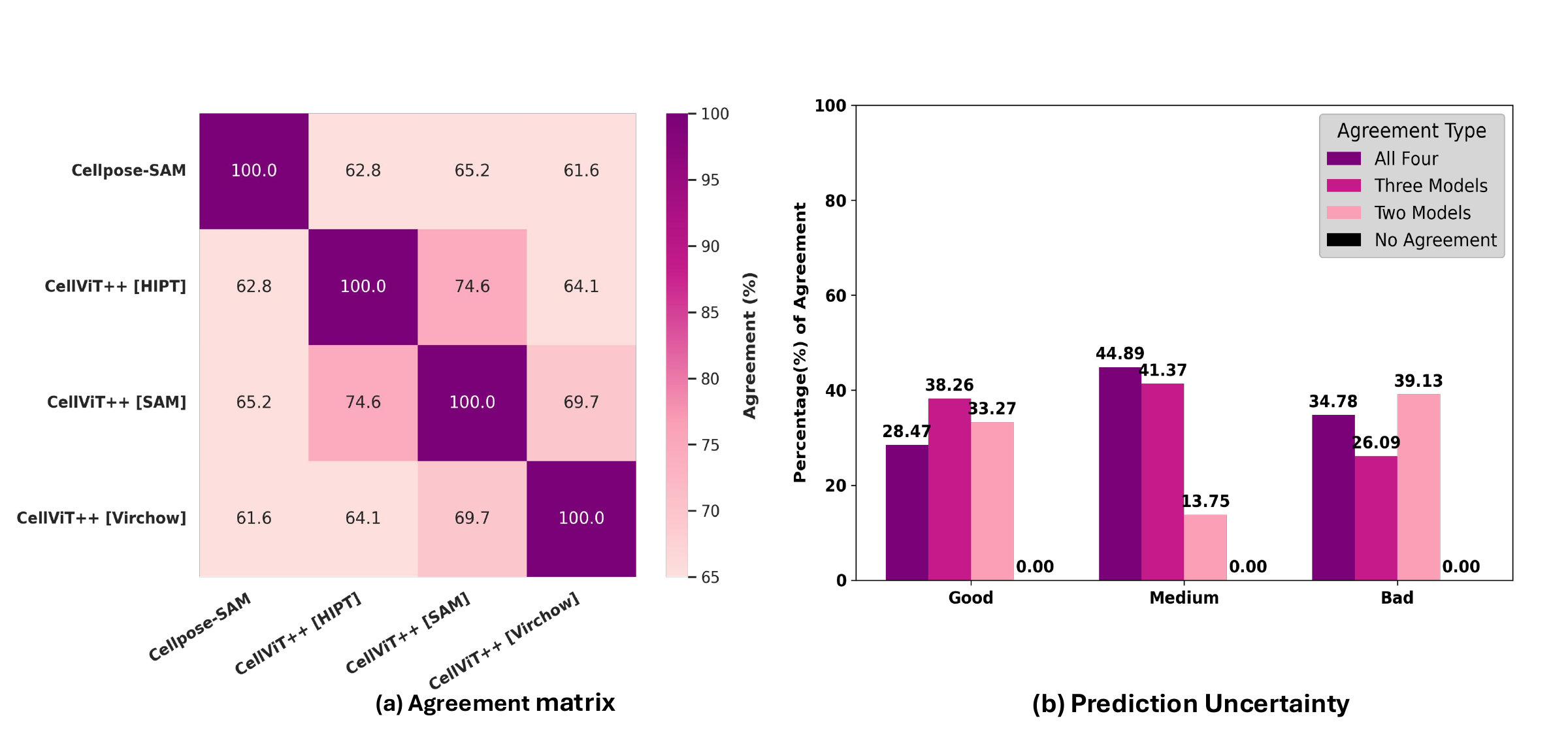} 
    \caption{\textbf{Cross-model performance rating agreement.}(a) shows the agreement percentages between each pair of foundation models used in this study. To further assess the cross-model performance, (b) shows the percentages of image patches where all four models agree, three agree, two agree, or no agreement, for each rating category (“Good”, “Medium”, “Bad”).}
    \label{fig:agreement-matrix}
\end{figure}

\noindent As shown in Fig.~\ref{fig:agreement-matrix}a, pairwise agreement scores among different models ranged from 61.6\% to 74.6\%, with the highest between CellViT++ [SAM] and CellViT++ [HIPT], reflecting shared instance segmentation framework and fine-tuning strategy. In contrast, Cellpose-SAM shows lower alignment with CellViT++ variants, reflecting architectural divergence and a distinct training regime.

To further dissect prediction uncertainty, we analyzed agreement per rating category. Fig.~\ref{fig:agreement-matrix}b shows that ``Medium" ratings had the highest 4-model consensus, while ``Good" had the lowest (28.47\% full agreement), indicating a larger performance gap among models on certain challenging image patches. ``Bad" predictions showed moderate agreement. To sum up, these results highlight meaningful prediction diversity across new AI cell foundation models and motivate our use of fusion strategies that combine multiple state-of-the-art cell foundation models to reduce prediction uncertainty and improve nuclei segmentation outcomes.



\subsection{Comparative Study with Previous Work. }

\begin{figure}[!h]
    \centering
    \includegraphics[width=1.0\textwidth]{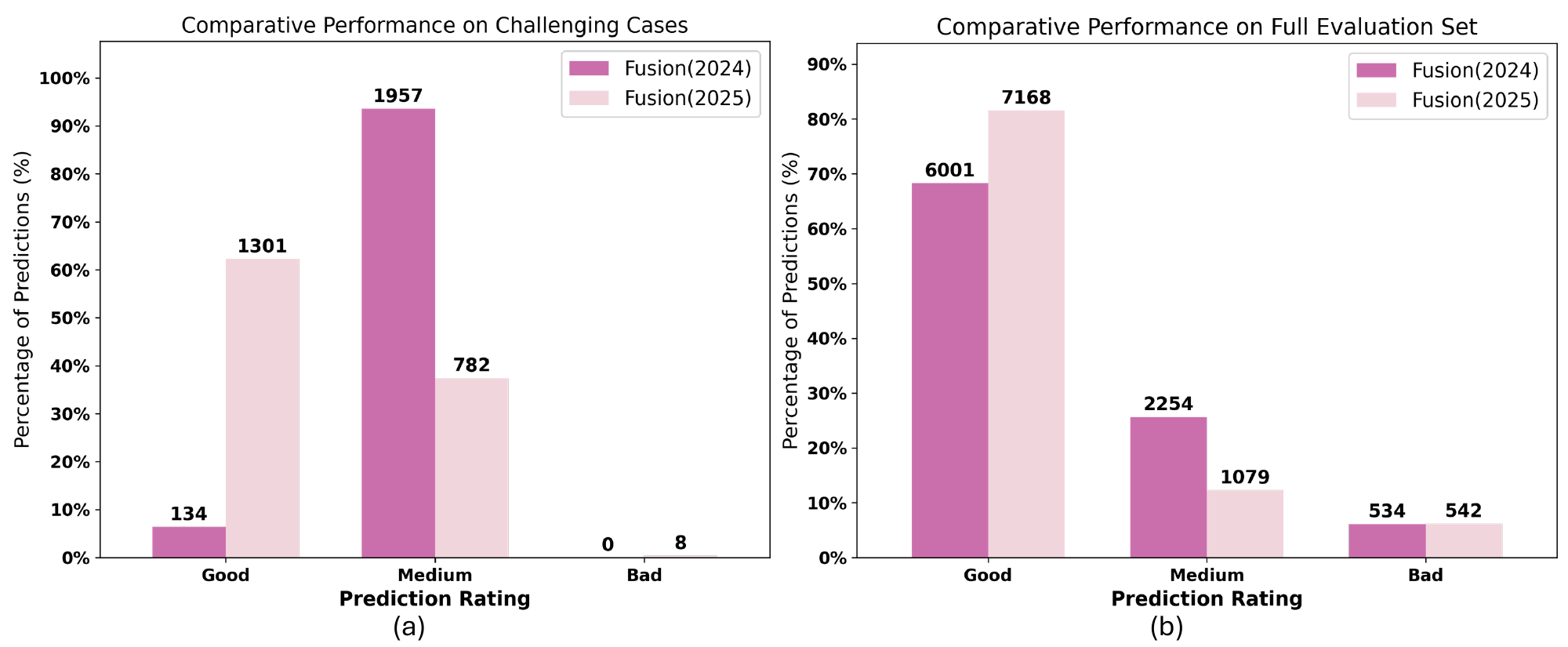}
    \caption{\textbf{Comparative analysis of fusion model performance with previous work.} (a) Evaluation of 2024 and 2025 fusion models on 2,091 challenging image patches. (b) Performance comparison on the full dataset (8,789 image patches). In both evaluations, the 2025 fusion model yields more ``Good" predictions, fewer ``Medium" predictions, while the proportion of ``Bad" remains stable.}
    \label{fig:comparasion}
\end{figure}

We evaluated the performance of our 2024 and 2025 fusion models on both challenging cases (2,091 image patches) and the complete evaluation dataset (8,789 image patches) to assess segmentation quality improvements. As shown in Fig.~\ref{fig:comparasion}a, the 2025 model significantly improved performance on challenging cases: ``Good” predictions rose from 134 (6.4\%) to 1,301 (62.2\%), ``Medium” dropped from 1,957 (93.5\%) to 782 (37.4\%). Fig.~\ref{fig:comparasion}b shows consistent gains on the full dataset, with ``Good” increasing from 6,001 (68.3\%) to 7,168 (81.6\%), ``Medium” decreasing from 2,254 (25.6\%) to 1,079 (12.3\%), and ``Bad” remaining stable at about 6.2\%. 

These results demonstrate that fusing updated AI cell foundation models improves nuclei instance segmentation precision and robustness, with particular strength in converting previously difficult samples to high-quality segmentation outputs. The substantial reduction in Medium-quality predictions across both evaluation scenarios validates the utility of this fusion strategy, based on multiple new AI cell foundation models, for automated renal histopathology analysis.





\subsection{Qualitative Results. }
\begin{figure}[!h]
    \centering
    \includegraphics[width=1\textwidth]{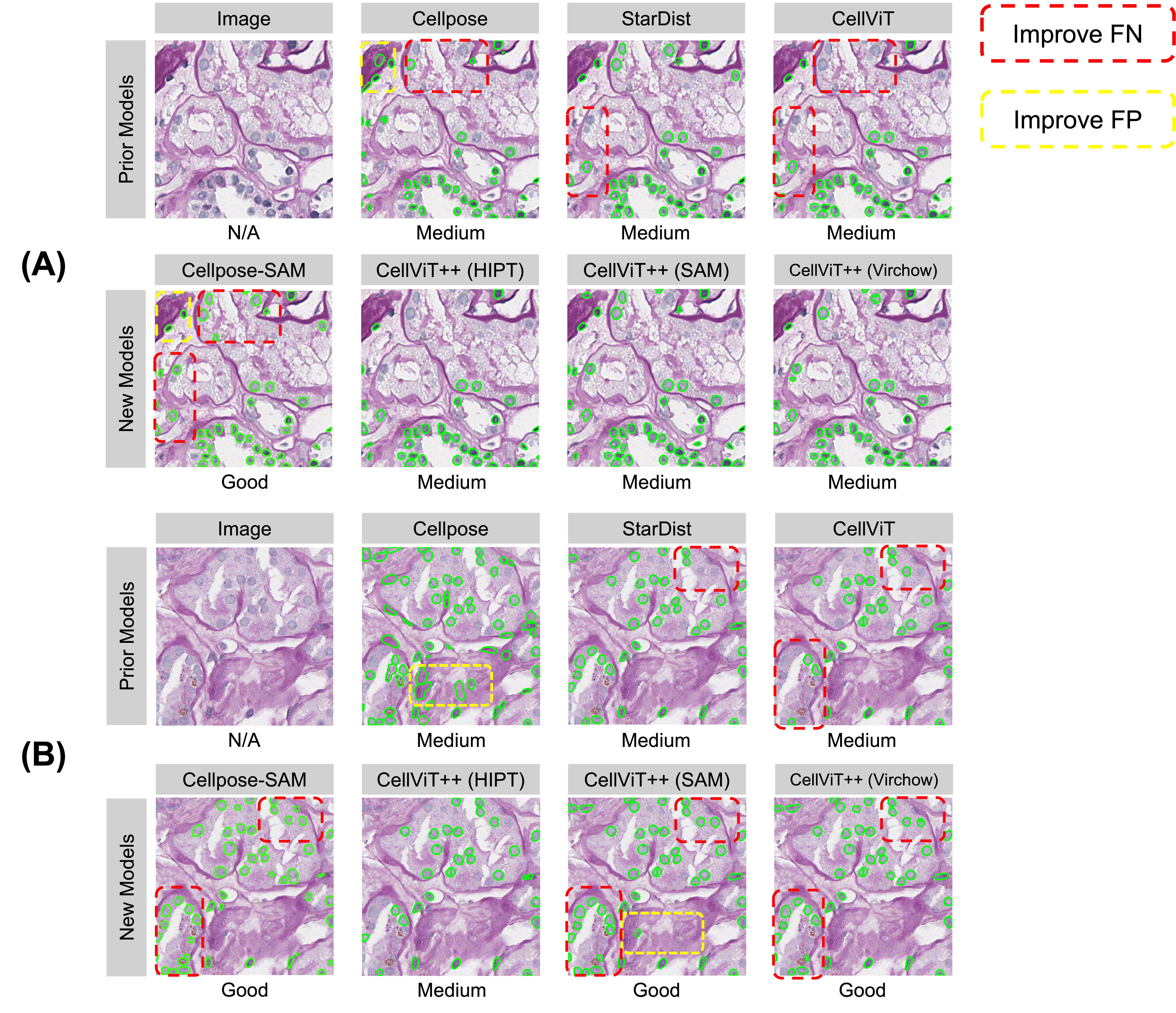}
    \caption{\textbf{Qualitative results.} Examples of patches where new AI cell foundation models produced ``Good" predictions missed by earlier models. Highlighted regions illustrate improvements in false positives (\textbf{FP}) and false negatives (\textbf{FN}) compared to prior AI cell foundation models: The \textbf{\textcolor{yellow}{yellow box}} marks areas where the new models reduce FP from the original predictions, while the \textbf{\textcolor{red}{red box}} highlights improved segmentation of FN regions.}
    \label{fig:qualitative_results}
\end{figure}

Fig.~\ref{fig:qualitative_results} shows two example patches where new AI cell foundation models produced ``Good" predictions missed by prior cell foundation models (Cellpose, StarDist, CellViT). In general, as shown in the original model predictions (first row in both examples), these image patches are challenging, with faint nuclei and low image contrast between the nuclei and the background. In Fig.~\ref{fig:qualitative_results}A, we can see that Cellpose-SAM reduces false negatives (FN) from StarDist and CellViT. The highlighted yellow box shows that Cellpose-SAM can even detect nuclei in regions with heavy staining, thereby reducing the false positives (FP) from the Cellpose prediction. Similarly, in Fig.~\ref{fig:qualitative_results}B, the new AI foundation models demonstrate improved segmentation performance, as most faint nuclei in the lower left corner of the image patch are correctly segmented and the FP from the original Cellpose prediction is eliminated in the updated model predictions.

\section{Conclusion}

In this study, we build upon our prior evaluation of foundation models for kidney cell nuclei segmentation by addressing a key limitation: the failure to robustly segment hard image patches. We benchmark SOTA AI cell foundation models (2025), including three CellViT++ variants and Cellpose-SAM, and introduced a fusion strategy that integrates their strengths. Our results show that the 2025 fusion model substantially outperforms previous models across the full evaluation set and resolves the vast majority of hard cases previously left unaddressed. Specifically, the updated AI cell foundation models increased the proportion of ``Good” segmentation outcomes while nearly eliminating ``Bad” predictions on 2,091 difficult samples. These findings highlight the value of ensemble multiple new AI cell foundation models in enhancing generalization and robustness for organ-specific applications. The curated hard image patches benchmark and improved nuclei segmentation outcomes offer a practical foundation for future model refinement and deployment in renal pathology workflows.

\acknowledgements
\begin{flushleft}
This research was supported by NIH R01DK135597 (Huo), DoD HT9425-23-1-0003 (HCY), and KPMP Glue Grant. This work was also supported by Vanderbilt Seed Success Grant, Vanderbilt Discovery Grant, and VISE Seed Grant. This project was supported by The Leona M. and Harry B. Helmsley Charitable Trust grant G-1903-03793 and G-2103-05128. This research was also supported by NIH grants R01EB033385, R01DK132338, REB017230, R01MH125931, and NSF 2040462. We extend gratitude to NVIDIA for their support by means of the NVIDIA hardware grant. This work was also supported by NSF NAIRR Pilot Award NAIRR240055.

\end{flushleft}
\bibliography{report}
\bibliographystyle{spiebib}

\end{document}